\documentclass{svjour3}                     
\smartqed  
\usepackage{graphicx}
\usepackage{bm}
\def\lesssim{\ \raise.3ex\hbox{$<$}\kern-0.8em\lower.7ex\hbox{$\sim$}\ }
\def\gesim{\ \raise.3ex\hbox{$>$}\kern-0.8em\lower.7ex\hbox{$\sim$}\ }

\journalname{Journal of Low Temperature Physics - QFS2009}

\begin{document}

\title{Pseudogap in fermionic density of states in the BCS-BEC crossover
of atomic  Fermi gases
}


\author{S. Tsuchiya\and R. Watanabe\and Y. Ohashi
}


\institute{S. Tsuchiya\and R. Watanabe\and Y. Ohashi\at
              Department of Physics, Keio University, 3-14-1, Hiyoshi, Kohoku-ku, Yokohama 223-8522, Japan\\
			  %
              \email{tsuchiya@rk.phys.keio.ac.jp}           
			   \and
		   S. Tsuchiya\and Y. Ohashi\at 
		   CREST(JST), 4-1-8 Honcho, Saitama 332-0012, Japan\\
}

\date{Received: date / Accepted: date}

\maketitle

\begin{abstract}

 We study pseudogap behaviors of ultracold Fermi gases in the
 BCS-BEC crossover region. We calculate the density of states (DOS), as
 well as the single-particle spectral weight, above the superfluid
 transition temperature $T_{\rm c}$ including 
 pairing fluctuations within a $T$-matrix approximation. We find that
 DOS exhibits a pseudogap structure in the BCS-BEC crossover region,
 which is most remarkable near the unitarity limit. We determine the
 pseudogap temperature $T^*$ at which the pseudogap structure in DOS
 disappears. We also introduce another temperature $T^{**}$ at which the
 BCS-like double-peak structure disappears in 
 the spectral weight. While one finds $T^*>T^{**}$ in the BCS regime,
 $T^{**}$ becomes higher than $T^*$ in the crossover and BEC regime. 
 We also determine the pseudogap region in the phase diagram in terms of
 temperature and pairing interaction. 

\end{abstract}

\section{Introduction}
\label{intro}
Recently, the BCS-BEC crossover has been realized in ultracold Fermi
gases\cite{BCSBEC}. In this phenomenon, using a tunable pairing
interaction associated with a Feshbach resonance, one can study Fermi
superfluids from the weak-coupling BCS regime to the strong coupling BEC
regime in a unified manner. Because of this advantage, superfluid Fermi
gases would be also useful for the study of high-$T_{\rm c}$ cuprates
with a strong pairing interaction.
\par
In the under-doped regime of high-$T_{\rm c}$ cuprates, the so-called
pseudogap structure has been observed in the density of states
(DOS)\cite{Damascelli}. As the origin of the pseudogap, strong pairing
fluctuations has been proposed\cite{Janko,Rohe,Yanase,Perali}. 
However, because of the complexity of
this system due to strongly correlated electrons, other possibilities,
such as antiferromagnetic spin fluctuations and a hidden ordered state,
have been also discussed. Thus, to confirm the pairing fluctuation
scenario, another simple system only having superfluid fluctuations
would be useful. 
\par
The cold Fermi gas system meets this demand. It is much simpler than
high-$T_{\rm c}$ cuprates, and pairing fluctuations dominate over the
BCS-BEC crossover physics. Indeed, the pseudogap phenomenon in this
system has been recently
predicted\cite{Janko,Perali,Haussmann}. Although the $s$-wave pairing
symmetry of superfluid Fermi gas is different from the $d$-wave one in
high-$T_{\rm c}$ cuprates, we can still expect that the study of
pseudogap phenomenon in cold Fermi gases would be helpful in
understanding the under-doped regime of high-$T_{\rm c}$ cuprates. Since
a photoemission-type experiment has recently become possible in cold
Fermi gases\cite{Stewart}, observation of strong-coupling effects on
single-particle excitations is now possible within the current
technology. 
\par
In this paper, we investigate pseudogap behaviors of atomic Fermi gases
above the superfluid transition temperature $T_{\rm c}$. Including
pairing fluctuations within a $T$-matrix approximation, we calculate DOS
and single-particle spectral weight. we examine how pairing fluctuations
affect them over the entire BCS-BEC crossover region. We also determine
the pseudogap regime in the phase diagram in terms of temperature and
the strength of pairing interaction. 
\par
\section{Formalism}
\label{sec:1}

We consider a uniform two-component Fermi gas described by pseudospin
$\sigma=\uparrow,\downarrow$. For a broad Feshbach resonance (which all
the current experiments are using), it is known that one can safely
study the interesting BCS-BEC crossover physics by using the ordinary
BCS model\cite{BCSBEC}, given by 
\begin{equation}
H= \sum_{\bm p,\sigma}\xi_{\bm p}c_{\bm
 p\sigma}^\dagger c_{\bm p\sigma}-U\sum_{\bm q}\sum_{\bm p,\bm
 p^\prime}c_{\bm p+\bm q/2\uparrow}^\dagger c_{-\bm p+\bm
 q/2\downarrow}^\dagger c_{-\bm p^\prime+\bm q/2\downarrow}c_{\bm p^\prime+\bm q/2\uparrow}.
\label{eq.1}
\end{equation}
Here, $c_{\bm p\sigma}$ is an annihilation operator of a Fermi atom with
pseudospin $\sigma=\uparrow,\downarrow$. $\xi_{\bm p}\equiv
\varepsilon_{\bm p}-\mu=p^2/2m-\mu$ is the kinetic energy, measured from
the chemical potential $\mu$, where $m$ is an atomic mass. The pairing
interaction ($U>0$) is assumed to be tunable by a Feshbach resonance. In
cold atom physics, the strength of pairing interaction is conveniently
described in terms of the parameter $(k_Fa_s)^{-1}$, where $a_s$ is the
$s$-wave scattering length and $k_F$ the Fermi momentum. 
In this scale, the BCS limit and BEC limit are, respectively, given by $(k_{\rm
F}a_s)^{-1}\ll-1$ and $(k_{\rm F}a_s)^{-1}\gg +1$.  
The region $-1\lesssim (k_{\rm F}a_s)^{-1}\lesssim
+1$ is referred to as the crossover region. The relation between $U$ and
$a_s$ is given by\cite{Randeria} $4\pi a_s/m=-U/[1-U\sum_{\bm
p}(m/p^2)]$. 
\par
The single-particle thermal Green's function is given by
$G_{\bm p}(i\omega_n)=1/((G^0_{\bm p}(i\omega_n))^{-1}-\Sigma(\bm p,i\omega_n))$.
Here, $\omega_n$ is the fermion Matsubara frequency, and $G^0_{\bm
p}(i\omega_n)=1/(i\omega_n-\xi_{\bm p})$ is the non-interacting Fermi Green's
function. The self-energy
$\Sigma(\bm p,i\omega_n)$ involves effects of pairing fluctuations. In
this paper, we include strong-coupling corrections within the $T$-matrix
approximation\cite{Janko,Rohe,Perali}. The resulting self-energy has the
form
\begin{equation}
\Sigma(\bm p,i\omega_n)=T\sum_{\bm q,\nu_n}\Gamma(\bm q,i\nu_n)G_{\bm
 q-\bm p}^0(i\nu_n-i\omega_n)e^{i(\nu_n-\omega_n)\delta},
\label{eq.3}
\end{equation}
where $\nu_n$ is the boson Matsubara frequency. 
The particle-particle scattering matrix
$\Gamma$ (describing pairing fluctuations) is given by $\Gamma({\bm
q},i\nu_n)=-U/[1-U\Pi(\bm q,i\nu_n)]$, where $\Pi(\bm
q,i\nu_n)=T\sum_{\bm p,\omega_n}G^0_{{\bm p}+{\bm
q}/2}(i\nu_n+i\omega_n)G^0_{{-\bm p}+{\bm q/2}}(-i\omega_n)$ is a
pair-propagator. 

\begin{figure*}
\includegraphics[width=\textwidth]{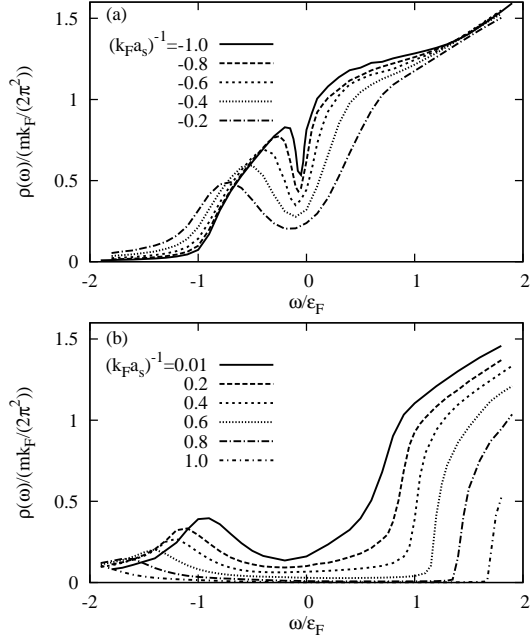}
\caption{Density of states $\rho(\omega)$ at $T_c$. (a) BCS side
 ($(k_{\rm F})^{-1}<0$). (b) BEC side ($(k_{\rm F})^{-1}>0$).} 
\label{dosTc}       
\end{figure*}

\par
To discuss the pseudogap phenomenon above $T_{\rm c}$, we need to
determine $T_{\rm c}$. Following Ref.\cite{NSR}, we employ the Thouless
criterion\cite{Thouless}, $\Gamma(\bm q=0,i\nu_n=0,T=T_{\rm c})^{-1}=0$,
and solve this equation, together with the equation for the number of
fermions, 
\begin{equation}
N=2T\sum_{\bm p,\omega_n}e^{i\omega_n\delta}G_{\bm p}(i\omega_n).
\label{eq.5}
\end{equation}
The above treatment can describe the smooth crossover behavior of
$T_{\rm c}$ and $\mu$ in the BCS-BEC crossover\cite{Perali,NSR}. Namely,
starting from the weak-coupling BCS regime, $T_{\rm c}$ gradually
deviates from the mean-field result to approach $T_{\rm
c}=0.218\varepsilon_{\rm F}$ of an $N/2$ ideal molecular Bose gas (where
$\varepsilon_{\rm F}$ is the Fermi energy). The chemical potential
monotonically decreases from $\varepsilon_{\rm F}$ in the crossover
regime to be negative in the BEC regime ($(k_Fa_s)^{-1}> 0.35$). The
negative $\mu$ indicates the formation of two-body bound states, so that
the BEC regime is well described by a molecular Bose gas, as expected.
\par
Above $T_{\rm c}$, we solve Eq. (\ref{eq.5}) to determine $\mu$. DOS
$\rho(\omega)$ and the spectral weight $A(\bm p,\omega)$ are,
respectively, evaluated from the analytic continued Green's function, as
\begin{eqnarray}
\rho(\omega)&=&-{1 \over \pi}\sum_{\bm p}{\rm Im}G_{\bm p}(i\omega_n\to\omega_+),
\label{DOS}
\\
A(\bm p,\omega)&=&-\frac{1}{\pi}G_{\bm p}(i\omega_n\to\omega_+).
\label{spectralf}
\end{eqnarray}
\par

\begin{figure*}
\includegraphics[width=\textwidth]{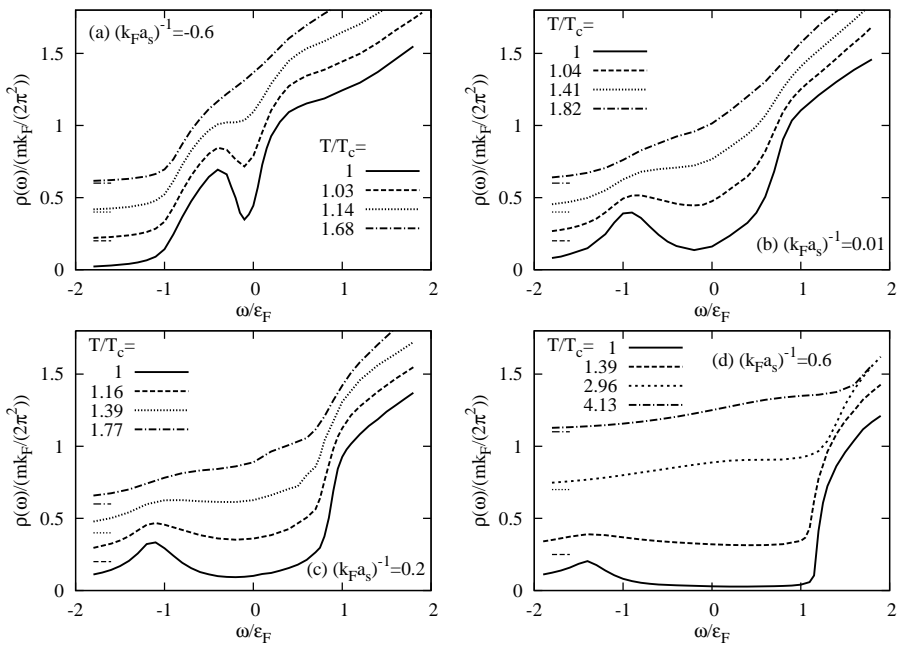}
\caption{Temperature dependence of the density of states
 $\rho(\omega)$. To clearly show how the pseudogap disappears, we have
 offset results for $T>T_{\rm c}$. The horizontal line in the left end
 of each curve indicates zero density of states.}
\label{dos} 
\end{figure*}

\section{Results}
\label{sec:2}

Figure \ref{dosTc} shows DOS at $T_{\rm c}$. In the BCS side ($(k_{\rm
F}a_s)^{-1}<0$) (panel (a)), we find a pseudogap (dip) structure around
$\omega=0$, which evolves as one approaches the
unitarity limit ($(k_{\rm F}a_s)^{-1}=0$). Since we only include pairing
fluctuations, this pseudogap purely originates from pre-formed pairs in
the normal state.
\par
In the BEC regime (where $\mu<0$), when we only include the negative $\mu$
and ignore other strong coupling effects, DOS has a finite gap as
$\rho(\omega)\propto\sqrt{\omega+|\mu|}$. Indeed, Fig.\ref{dosTc}(b)
shows that DOS continuously changes into the {\it fully} gapped
structure. Since $2|\mu|$ equals the binding energy of a molecule in the
BEC limit\cite{NSR}, the almost fully gapped structure at $(k_{\rm
F}a_s)^{-1}=1$ in Fig.\ref{dosTc}(b) indicates that the system is rather
close to a molecular Bose gas than a Fermi atom gas.  

\begin{figure*}
\begin{center}
\includegraphics[width=0.65\textwidth]{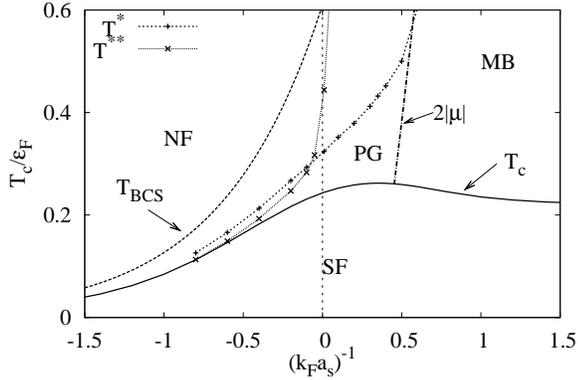}
\end{center}
\caption{Pseudogap temperatures $T^*$ and $T^{**}$ determined from DOS
 and spectral weight, respectively. In this figure, SF is the superfluid
 phase, PG is the pseudogap regime, and MB is the molecular Bose gas
 regime. Above $T^*$ or $T^{**}$, the system is regarded as a normal
 Fermi gas (NF) with weak pairing fluctuations.}
\label{phdgm}       
\end{figure*}

Figure \ref{dos} shows DOS above $T_{\rm c}$. As expected, the pseudogap
gradually disappears as one increases the temperature. When we define
the pseudogap temperature $T^*$ at which the dip structure in DOS
disappears, we obtain the phase diagram Fig.\ref{phdgm}. Although $T_{\rm c}$ calculated
in the mean-field theory is usually considered as a characteristic
temperature at which preformed pairs appear, Fig.\ref{phdgm} shows
that the pseudogap temperature $T^*$ evaluated from DOS is actually much
lower than the mean-field value of $T_{\rm c}$ ($T_{BCS}$ in this figure).

\begin{figure*}
\includegraphics[width=\textwidth]{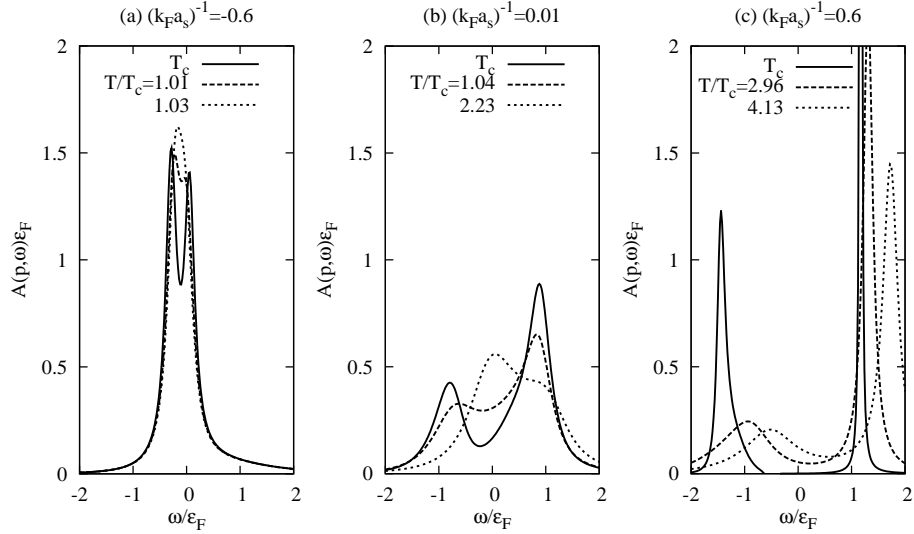}
\caption{Spectral weight $A(\bm p,\omega)$ as a function of $\omega$. In each panel, we take the momentum where the peak-to-peak energy is minimum: (a) $p/k_F=0.91$, (b) 0.83, and (c) 0.01.}
\label{sw}       
\end{figure*}

In the BCS theory, the spectral weight below $T_{\rm c}$ has the double-peak structure as 
\begin{eqnarray}
A({\bm p},\omega)=
{1 \over 2}
\Bigl[
1+{\xi_{\bm p} \over E_{\bm p}}
\Bigr]
\delta(\omega-E_{\bm p})
+
{1 \over 2}
\Bigl[
1-{\xi_{\bm p} \over E_{\bm p}}
\Bigr]
\delta(\omega+E_{\bm p}).
\label{eq.6}
\end{eqnarray}
Here, $E_{\bm p}=\sqrt{\xi_{\bm p}^2+\Delta^2}$ is the Bogoliubov
excitation spectrum, where $\Delta$ is the superfluid order
parameter. The minimum value of peak-to-peak energy equals $2\Delta$. In
the simple mean-field theory, this double-peak structure only exists
below $T_{\rm c}$. However, when one includes pairing fluctuations, the
double-peak structure still remains above $T_{\rm
c}$\cite{Janko,Perali}, as shown in Fig.\ref{sw}. This characteristic
structure disappears at a certain temperature $T^{**}$ (See
Fig.\ref{sw}(a) and (b)), which we can define as another pseudogap temperature.  

When we plot $T^{**}$ in Fig.\ref{phdgm}, we find that it does not
coincide with $T^*$. While the latter is higher in the BCS side, one
finds $T^{**}>T^*$ in the BEC side. In the BCS side, although the
double-peak structure is absent when $T\ge T^{**}$, pairing fluctuations still
strongly affect $A({\bm p},\omega)$ around the Fermi level
$p\sim\sqrt{2m\mu}$, leading to a broad and {\it low} single peak
structure. Then, since DOS is given by the momentum summation of $A({\bm
p},\omega)$, DOS around $\omega\sim 0$ (which is dominated by
$A(p\sim\sqrt{2m\mu},\omega)$) is suppressed, giving the dip
structure. On the other hand, as shown in Fig.\ref{sw}(c), the lower
peak in $A({\bm p},\omega)$ soon becomes broad above $T_{\rm c}$ in the
BEC regime, so that the effect of lower peak is easily smeared out in
the momentum summation in Eq. (\ref{eq.6}). As a result, DOS does not
reflect the double-peak structure in $A({\bm p},\omega)$ when $T^*<T\le
T^{**}$. 
\par
The above result indicates that the pseudogap temperature depends on
what we measure. When we measure a quantity where DOS is crucial, such
as the specific heat, $T^*$ would be the crossover temperature between
the pseudogap regime (PG) and normal Fermi gas regime (NF). On the other
hand, when we consider a quantity dominated by the spectral weight, such
as the photoemission-type experiment done by JILA group\cite{Stewart},
$T^{**}$ would work as the crossover temperature between PG and NF.

\par
As discussed previously, the pseudogapped DOS continuously changes into fully gapped one in the BEC regime, reflecting that the system reduces to an $N/2$ molecular Bose gas (MB). As a crossover temperature between PG and MB regime, the molecular binding energy $E_{\rm bind}$ would be useful, because thermal dissociation of molecules is suppressed when $T\lesssim E_{\rm bind}$. In the present case, noting that $E_{\rm bind}\simeq|2\mu|$ in the BEC regime (where $\mu<0$), one may conveniently determine the pseudogap regime as the region surrounded by $T_{\rm c}$, $2|\mu|$, and $T^{*}$ or $T^{**}$ in Fig.\ref{phdgm}.

\section{Summary}

To summarize, we have discussed effects of pairing fluctuations on
single-particle properties of a cold Fermi gas above $T_{\rm c}$. Within
the framework of $T$-matrix approximation, we showed how the pseudogap
appears in the density of states and the single-particle spectral weight
over the entire BCS-BEC crossover region. We have also determined the
pseudogap regime in the BCS-BEC crossover phase diagram. This phase
diagram would be useful for the observation of pseudogap phenomenon in
this system.

\begin{acknowledgements}
We thank A. Griffin for useful discussions.
\end{acknowledgements}


\end{document}